# Soliton pair dynamics in patterned ferromagnetic ellipses

Kristen S. Buchanan, [1] Pierre E. Roy, [2,1] Marcos Grimsditch,[1] Frank Y. Fradin, [1] Konstantin Yu. Guslienko, [1] Sam D. Bader, [1] and Valentyn Novosad[1]

[1]*Materials Science Division and Center for Nanoscale Materials, Argonne National Laboratory, Argonne, IL 60439*

[2]*Uppsala University, Department of Engineering Science, Uppsala, Sweden*

Confinement in nanomagnets alters their energetics and leads to new magnetic states, for example, vortices. There are many basic questions concerning dynamics and interaction effects that remain to be answered and nanomagnets are convenient model systems for studying these fundamental physical phenomena. A single vortex in restricted geometry, also known as a nonlocalized soliton, possesses a characteristic translational excitation mode that corresponds to spiral-like motion of the vortex core around its equilibrium position. Here we investigate the dynamics of magnetic soliton pairs confined in lithographically defined Permalloy ellipses using a microwave reflection technique. Strong resonances were detected experimentally in the sub-GHz frequency range and, by comparing with micromagnetic simulations, assigned to the translational modes of vortex pairs with parallel or antiparallel core polarizations. Although the vortex polarizations play a negligible role in the static interaction between two vortices, their effect dominates the dynamics.

In 1844[1] it was first recognized that the propagation of a solitary wave – known as a soliton - represents an unusual state of matter. Presently solitons have become ubiquitous in many fields of physics.[2,3,4,5,6,7,8] It is now understood that their origin can be traced to the nonlinearities in the physics that governs them. Static solitons exist in many forms: domain walls, vortices in ferromagnets, superconductors, and Bose-Einstein condensates, and in polymer chains. Some solitons exist only in dynamic form, such as, tsunamis and internal oceanic waves. The statics and dynamics of the interactions between solitons is a fascinating field: while tsunamis and domains can at times be described as almost non-interacting, they can also annihilate in pairs. In other situations, e.g., superconducting vortex lattices, there are strong interactions between the vortices that have a pronounced effect on their dynamics. Domain walls[9] and magnetic vortices[10] in geometrically confined ferromagnets provide model systems for the study of static and dynamic interactions between solitons (localized solutions of non-linear equations).

Here we present experimental evidence of strong dynamic interactions between two solitons (vortices) trapped in an elliptically-shaped ferromagnetic particle. This relatively simple system provides a unique opportunity to investigate the peculiarities of soliton - soliton interactions. The resonance frequencies were measured in microwave reflectivity experiments and associated, by comparing with micromagnetic modeling, with different combinations of in-phase and out-of-phase vortex core motion along the two ellipse axes. While the vortex polarizations have almost no effect on their static interaction, they play a dominant role in their dynamics.

A magnetic vortex consists of an in-plane flux-closure magnetization distribution with a very small central core that is magnetized perpendicular to the plane.[11] It was demonstrated by spin-polarized scanning tunneling microscopy that the core radius is ~10 nm, i.e., comparable to the material exchange length.[12] The vortex state can be characterized by a polarization $p = \pm 1$ that describes whether the core is oriented up or down, and a chirality that describes the direction of the in-plane curling of the magnetization. The magnetic vortex is an example of a nonlocalized rather than a localized soliton because the magnetization distribution still depends on position, even at large distances from the core.[8] Mesoscopic magnetically soft structures undergo magnetization reversal through the successive nucleation, displacement and annihilation of magnetic vortices.[13,14] A magnetic vortex core experiences a Magnus-type force (gyroforce) perpendicular to its velocity,[15,16] which is proportional to $p$. The gyroforce results in a relatively slow (sub-GHz frequency), spiral-like core motion,[17,18,19] an excitation known as a translational or gyrotropic mode. The handedness of the spiral depends only on the polarization, as demonstrated in time-resolved magneto-optical Kerr[20, 21, 22] and X-ray magnetic circular dichroism experiments.[23, 24] This is the only mode that is characteristic of a magnetic vortex. Other higher frequency (GHz range) excitations exist[25, 26] that can be understood in terms of quantized spin waves in restricted geometries.[27]

Vortex pair motion has been studied analytically based on Thiele's approach[15] and spin dynamics simulations for 2-dimensional anisotropic Heisenberg model[28, 29] This work shows that the core polarizations in an interacting vortex pair essentially influence their behavior. The case of vortex-anitivortex interactions has also been considered.[30] Furthermore, recent micromagnetic modeling predicts that magnetostatic interactions between patterned magnetic dots in close proximity, each containing a single vortex, will essentially alter the excitation spectrum.[31, 32] Thus the interactions between vortices confined in the same particle, e.g., a vortex pair in an elliptical magnetic dot, should also result in a modified spectrum. The static properties of this system have been investigated; however, no published reports exist concerning experimental observations of dynamic interactions between two magnetic vortices confined in a ferromagnet with restricted geometry.

Studies of the static properties of elliptical magnetic particles show that ellipses can be designed to support two magnetic vortices by tailoring the particle aspect ratio[33] or, as done in the present study, by controlling the nucleation mechanism.[34] Herein we investigate the excitations of two magnetic vortices confined in elliptical particles using a microwave reflection technique. A radio frequency (*r.f.*) current in a coplanar waveguide (CPW) generates an oscillating magnetic field that is absorbed preferentially when the frequency coincides with a resonance mode of the magnetic dots patterned on its central strip. Elliptically shaped Permalloy ($Fe_{20}$-$Ni_{80}$ alloy) dots with lateral design dimensions of 3 x 1.5 µm and thickness of 40 nm were patterned directly onto the waveguides by means of high resolution electron beam lithography, magnetron sputtering and lift-off techniques. There are ~2,000 particles on each waveguide. The dots are separated by >1 µm (edge-to-edge), therefore the magnetostatic interaction are small and can be neglected.

The microwave measurements were performed using a vector network analyzer operating in the reflection mode.[35] A static magnetic field $H$ is applied in-plane, either parallel or perpendicular to the *r.f.* excitation field. Samples were fabricated with the long axis either parallel or perpendicular to the CPW to explore the effects of excitation direction. The derivatives of the impedance (both real and imaginary parts) are recorded with respect to a small modulation field that is applied parallel to the static magnetic field.

Magnetic force microscopy (MFM) imaging, sensitive to the stray magnetic field gradient, confirms that after saturating the sample along the long (easy) axis (*x*), the majority of the ellipses are in a two-vortex state at remanence (Fig. 1a), whereas saturation along the short (hard) axis (*y*) results in a single-vortex remanent state. This strong correlation between saturation direction and remanent state is consistent with an earlier report on the remanent states of magnetically soft ellipses of similar size.[34] The in-plane magnetization distribution and core profiles associated with the static two-vortex state are shown in Figs. 1b and c, respectively, as calculated using micromagnetic simulations based on the Landau Lifshitz-Gilbert equation [Scheinfein, M. R. LLG Micromagnetics Simulator,™ http://llgmicro.home.mindspring.com/]. Two vortices confined in an ellipse will necessarily have opposing chiralities but may have cores with the same or opposite polarizations, i.e., $p_1 = p_2$ or $p_1 = -p_2$. Note that the simulations lead to the same energies and core positions independent of the polarizations.

Figure 2a shows representative microwave impedance derivatives, real and imaginary, as a function of frequency where the *r.f.* field is oriented along the short axis and a static field of $H = 10$ Oe is applied along the long axis. The resonance peaks are defined by taking the average of the peak position on the real spectrum with the zero crossing of the imaginary spectrum. This procedure introduces at most a ±5 MHz systematic error in the resonance position. These spectra were obtained after first saturating the ellipses along the long axis, i.e., they are associated with a vortex pair. When the sample is saturated along the short axis (single vortex state) the two peaks virtually disappear and are replaced by a single, lower frequency peak. We now focus on the novel excitations associated with the two-vortex state. The properties of the single vortex state in ellipses will be presented elsewhere.

The field dependence of the detected resonance lines are summarized in Fig. 2b. In this figure, the results for *H* applied along the long and short axes are displayed at negative and positive fields, respectively. The observed eigenmodes, in both cases, are symmetric with respect to the sign of the applied field. Solid (open) symbols indicated that *H* is perpendicular (parallel) to the *r.f.* field. When *H* is applied along the long axis the peaks are only observed over a small field range because the vortex annihilation field is much lower for this configuration. The frequencies of the two observed peaks decrease slightly with increasing *H*. Near zero field, their frequencies are 188 and 142 MHz. The modes marked by up-triangles (~142 MHz) are observed independent of the perturbation field direction. This resonance mode splits into two branches as the field along the short axis is increased - one branch increases in frequency while the other decreases. In contrast, those marked by down-triangles are associated with a particular excitation direction. A perturbation along *y* excites the high frequency (~188 MHz) mode that changes little with *H*. At fields above 40 Oe an along *y* an additional line was detected at a lower frequency (80 MHz at 40 Oe) for a perturbation along *x*. This mode increases in frequency with *H* and intersects the effective low-frequency cut-off of the instrumentation (~70 MHz). These results are quite different from those reported for single vortices in cylindrical dots where the vortex translational mode frequency is virtually independent of *H*.[35]

To gain insight into the origins of the observed excitations, micromagnetic modelling was conducted. The simulations are based on the numerical solution of the Landau-Lifshitz-Gilbert equation, an equation that describes the time-evolution of a magnetic moment in an effective magnetic field that includes contributions from magnetic exchange, dipolar fields, and external magnetic fields. A magnetic structure is represented by an array of rectangular cells, each with a magnetic moment of constant magnitude. Dynamic simulations of the translational-mode dynamics were carried out by first allowing a particular remanent state to relax to equilibrium under the influence of a small, in-plane perturbation field using a large damping parameter ($\alpha = 1.0$). Then, the perturbation field is removed and the simulation is continued using a smaller damping parameter ($\alpha = 0.008$). Similar to the case of a single vortex in a cylindrical dot, we find that the vortex cores rotate in elliptical orbits around their equilibrium positions. The frequency can be extracted from the temporal oscillations of the average magnetization or by tracking the core position. In order to achieve reasonable simulation times (~12 hour runs on a desktop computer), the simulations were conducted for a 1.5 by 0.75 µm ellipse, 20 nm thick, exactly half the size of the experimental sample, with an in-plane cell size of $5 \times 5$ nm$^2$. Magnetic parameters appropriate for bulk Permalloy with negligible magnetic anisotropy were used: exchange constant $A = 1.3$ µerg/cm, saturation magnetization $M_s = 800$ emu/cm$^3$, and gyromagnetic ratio $\gamma = 18.5$ MHz/Oe. Simulations were conducted for both vortex polarization combinations $p_1 p_2 = +1$ and $p_1 p_2 = -1$. Perturbation fields were applied along *x* and *y* (the ellipse major and minor axes, respectively).

The micromagnetic simulations show that different modes, each with a unique eigenfrequency, are excited depending on the core polarizations and the *r.f.* field direction. For all modes, each vortex core circulates around its equilibrium position at the same frequency. Although neither the core equilibrium positions nor the total energy of the particle is affected by the relative core polarizations, the eigenfrequencies for the two $p_1 p_2 =$

±1 cases are substantially different. The reason for this difference stems from the fact that the sign of the polarization determines the sense of rotation. Figure 3a is a schematic of the vortex core motion for the conceivable modes for $p_1 p_2 = \pm 1$. The modes can be differentiated based on the phase between the motions of the two cores. When $p_1 p_2 = +1$, both cores rotate in the same direction so that the two possible modes are characterized by either in-phase *(i)* or out-of-phase *(o)* motion of the cores along the *(x,y)* directions. When $p_1 p_2 = -1$ the cores rotate in opposite directions so it is no longer possible to label the modes as *(i, i)* or *(o, o)*; instead we label them as *(i, o)* and *(o, i)*, respectively. It is this distinction that is the origin of the different behaviors for $p_1 p_2 = \pm 1$.

Figure 3b illustrates how the core equilibrium positions change in the presence of a magnetic field. In general a static field will move the vortex core equilibrium positions, resulting in different restoring forces and frequencies. When a magnetic field is applied along *x*, the two vortices shift in opposite directions by equal amounts along *y*, resulting in states that are equivalent upon field inversion. When the field is applied along *y*, however, the cores will separate when the magnetization of the central region $m_c$ is parallel to *H* and move together when $m_c \cdot H < 0$ (see Fig. 1b). For a given field, these two states are not equivalent and hence their frequencies will not necessarily be the same. Because there are many ellipses on the stripline, we expect equal numbers with $m_c$ oriented parallel or antiparallel to *y*. Thus the frequency 'splitting' of the modes observed when a field is applied along the short axis can be understood as being due to a mixture of the two vortex chiralities.

In light of the core motions shown in Fig. 3a, it is clear that a uniform external field can only excite modes with out-of-phase core motion. For $p_1 p_2 = +1$ the *(o, o)* (Supplementary Video 1) mode will be excited by either direction of the perturbing field, while the *(i, i)* mode will never be excited. For the $p_1 p_2 = -1$ case, however, a field along *x* excites the *(i, o)* mode (Supplementary Video 2) while a field along *y* excites the *(o, i)* mode (Supplementary Video 3).

Figure 4 shows a plot of the calculated frequencies for the different polarization combinations and excitation directions. We denote simulations performed for perturbation fields perpendicular to the static field by full symbols and full lines, and those for parallel driving fields by open symbols and dashed lines. Since for $p_1 p_2 = +1$ the same mode is excited by both fields, the results overlap. The open triangle with an "x" through it with a frequency of 138 MHz at $H = 0$ was obtained from a micromagnetic simulation of the in-phase $p_1 p_2 = +1$ mode, the mode illustrated in the upper left of Fig. 3a. It was excited by displacing the cores symmetrically in a particle with $p_1 p_2 = +1$, a perturbation that cannot be induced by a uniform external field. The existence of this mode however does confirm the assumptions that led us to postulate the modes in Fig. 3a.

The modes in Figure 4 provide an excellent qualitative description of the experimental results in Fig. 2b. For $p_1 p_2 = +1$ the same mode is excited for both perturbation fields and the 'splitting' observed for $H$ along $y$ is due to the presence of a population of particles with chiralities that have the magnetization of the central region both aligned with and against the magnetic field. For $p_1 p_2 = -1$, we observe the mode excited by a perturbation along $y$ experimentally but do not detect the small mode splitting for $H$ along $y$. We observe only the upper frequency mode of the split pair for a perturbation along $x$; the lower branch and the resonance for $H$ along $x$ are unresolved in this experiment. The frequencies in the simulations are uniformly higher than those observed in the experiment. Tests for selected cases indicate that the scaling introduces an error of up to 10%. The simulations are also sensitive to the ellipse dimensions so deviations of the sample size, shape and $M_s$ from the model will also contribute to the discrepancy.

Earlier theoretical works based on Thiele's approach[15] applied to the anisotropic Heisenberg spin model predicted that the core polarizations in an interacting vortex pair do influence their behavior.[28,29] These theories, however, do not account for the applied magnetic field or the magnetostatic energy contributions and therefore have limited capacity to predict the eigenfrequencies and core trajectories for our case. To gain insight into the coupled vortex dynamics we have also solved a system of Thiele's equations[15] $\mathbf{G}_j \times \frac{d\mathbf{X}_j}{dt} - \frac{\partial W(\mathbf{X}_1, \mathbf{X}_2)}{\partial \mathbf{X}_j} = 0$ but with a phenomenological energy function $W(\mathbf{X}_1, \mathbf{X}_2)$, where $\mathbf{X}_j = (X_j, Y_j)$ are the core positions of vortices $j = 1, 2$. The energy function $W(\mathbf{X}_1, \mathbf{X}_2)$ that is responsible for the restoring force (second term) contains contributions from the self-induced magnetostatic charges and exchange interactions between spins but magnetostatic contributions dominate for micron-sized dots. The first term, the gyroforce, is proportional to the gyrovector $\mathbf{G}_j = -Gp_j\hat{\mathbf{z}}$, where $G = 2\pi LM_s/\gamma$, $\gamma$ is the gyromagnetic ratio, $M_s$ is the saturation magnetization and $\hat{\mathbf{z}}$ is the unit vector perpendicular to the dot plane. The stiffness coefficients related to the restoring forces can be defined as the derivatives $\kappa_{ij}^{\alpha\beta} = \partial^2 W(\mathbf{X}_1, \mathbf{X}_2)/\partial X_i^\alpha \partial X_j^\beta$ evaluated at the equillbrium positions $X_{jo}$ with $\alpha, \beta = x, y$. Due to the elliptic symmetry of the dot, the only nonzero components are $\kappa_{ii}^{\alpha\alpha} = \kappa_\alpha$ and $\kappa_{12}^{\alpha\alpha} = \kappa_{21}^{\alpha\alpha} = \mu_\alpha$; the latter are due to vortex-vortex interactions. Solving the system of coupled linear equations one finds the eigenfrequencies $\omega_{1,2}^{+} = G^{-1}\sqrt{(\kappa_x \pm \mu_x)(\kappa_y \pm \mu_y)}$ for $p_1 p_2 = +1$, and $\omega_{1,2}^{-} = G^{-1}\sqrt{(\kappa_x \pm \mu_x)(\kappa_y \mp \mu_y)}$ for $p_1 p_2 = -1$. The eigenvectors of these equations provide information regarding the phases between the vortex core displacements and yield the same phase relationships (o, o), (i, o), (o, i), and (i, i) illustrated in Fig. 3 for the different modes. In the experiment, a sinusoidal magnetic field perturbs the vortex cores from their equilibrium positions in the elliptical dot. Their subsequent motion in dynamical potential wells near their equilibrium positions is governed by the gyroforce and dynamical magnetostatic forces and is detected as a change in impedance of the waveguide at a resonance frequency.

One of the properties of the above eigenfrequencies is that the product of the two frequencies for each polarization combination should be invariant; *viz.* $\omega_1^{+}\omega_2^{+} = \omega_1^{-}\omega_2^{-}$. This conjecture is confirmed by the micromagnetic simulation results shown in Fig. 4 where the frequency products for $p_1 p_2 = +1$ and $p_1 p_2 = -1$ are (2.79 ± 0.03) x$10^4$ MHz$^2$ and (2.79 ± 0.03) x$10^4$ MHz$^2$, respectively, assuming a 1 MHz uncertainty in frequency (typical variation in frequency depending on the number of oscillations used in the fit). The agreement is excellent, suggesting that the phenomenological model provides a good description of the vortex-pair resonance behavior and may serve as a guide to detailed theoretical formulations of the problem. Although not presented here, we have also found that the invariance of the frequency product continues to hold at non-zero field values.

In conclusion, the static and dynamic interactions between vortices are very different; the static properties are essentially independent of polarization, while the dynamic behavior is dominated by the polarization of the vortices. Using a broadband microwave reflection technique we have detected and identified the translational eigenmodes characteristic of coupled pairs of magnetic vortices confined within an elliptical dot. Four modes were identified from the micromagnetic simulations corresponding to different combinations of in-phase and out-of-phase core motion along the two ellipse axes, three of which can be excited by a spatially uniform perturbation field. One observed mode corresponds to identical core polarizations and is independent of excitation direction, while the other two observable modes correspond to in-phase-x and out-of-phase-x excitations of vortex pairs with opposing polarization.

Correspondence and requests for materials should be addressed to V. Novosad (novosad@anl.gov).

**Acknowledgements**

 We thank Y. Otani and J. Pearson for stimulating discussions and R. Divan for lithography support.  This work was supported by the U.S. Department of Energy, Basic Energy Sciences, Material Sciences under Contract No. W-31-109-ENG-38. K. B. thanks NSERC of Canada for a fellowship. P.R. acknowledges support from the Swedish Research Council and Swedish Foundation for Strategic Research.

**Figure captions**

**Figure 1**: MFM and simulations of the two-vortex remanent states. a) MFM images of a 3 by 1.5 μm ellipse, 40 nm thick. Half-size simulations (1.5 by 0.75 μm ellipse, 20 nm thick) show that the in-plane equilibrium magnetization distribution b) and core positions c) are independent of the core polarizations.

**Figure 2:** Experimentally measured vortex-pair resonance frequencies. a) Representative real (solid) and imaginary (dotted) impedance derivative spectra for *H* of 10 Oe applied along the long axis of the 3 by 1.5 μm Permalloy ellipses shown in the optical microscope image (inset). Two peaks are observed after first saturating the sample along the long axis. The field dependence of the resonance frequencies is shown in b), where the results for *H* along the long and short axes are displayed at negative and positive fields, respectively. Solid and open symbols indicate that the r.f. field was applied perpendicular or parallel to *H*, respectively.

**Figure 3**: Diagrams of the magnetization configurations and dynamic modes for two magnetic vortices confined in an ellipse. a) The vortex translational modes observed in micromagnetic simulations are shown, where the solid and open dots represent the two possible core polarizations and "x" marks the core equilibrium positions. b) When a static field is applied along the *y*-axis, the cores shift either together or apart depending on whether the central region is magnetized antiparallel or parallel to *H*. A field applied along *x*, in contrast, results in equivalent states. The greyscale represents the *y*-component of the magnetization, where white represents positive saturation.

**Figure 4**: Micromagnetic simulation results showing the excitation frequencies as a function of *H*. The simulations are for two vortices confined in a 1.5 by 0.75 μm ellipse, 20 nm thick. Solid and open symbols show eigenmodes excited by an r.f. field perpendicular or parallel to *H*, respectively. The triangle with an "x" through it at zero field represents a mode that is not excited by this experiment. Diagrams indicate the direction of *H* and the core positions for the branches that result from applying *H* parallel or antiparallel to the central magnetization.

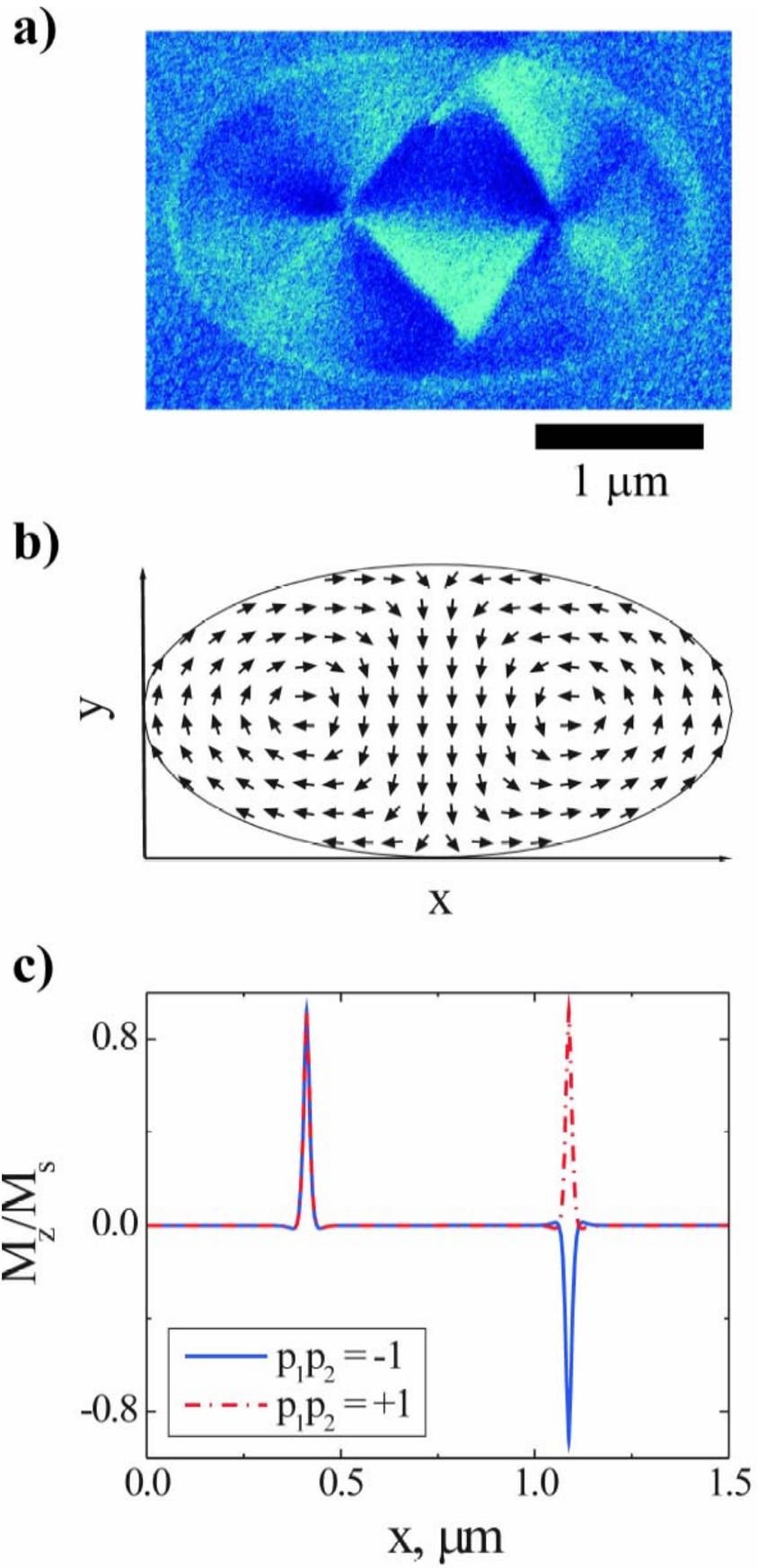

Figure 1

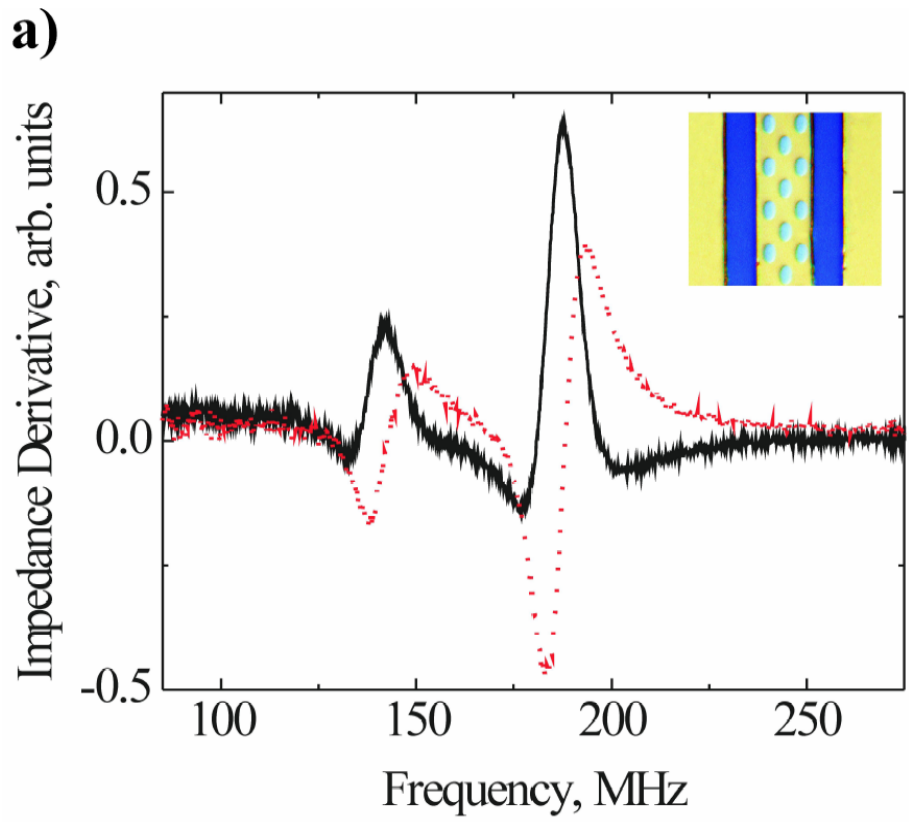

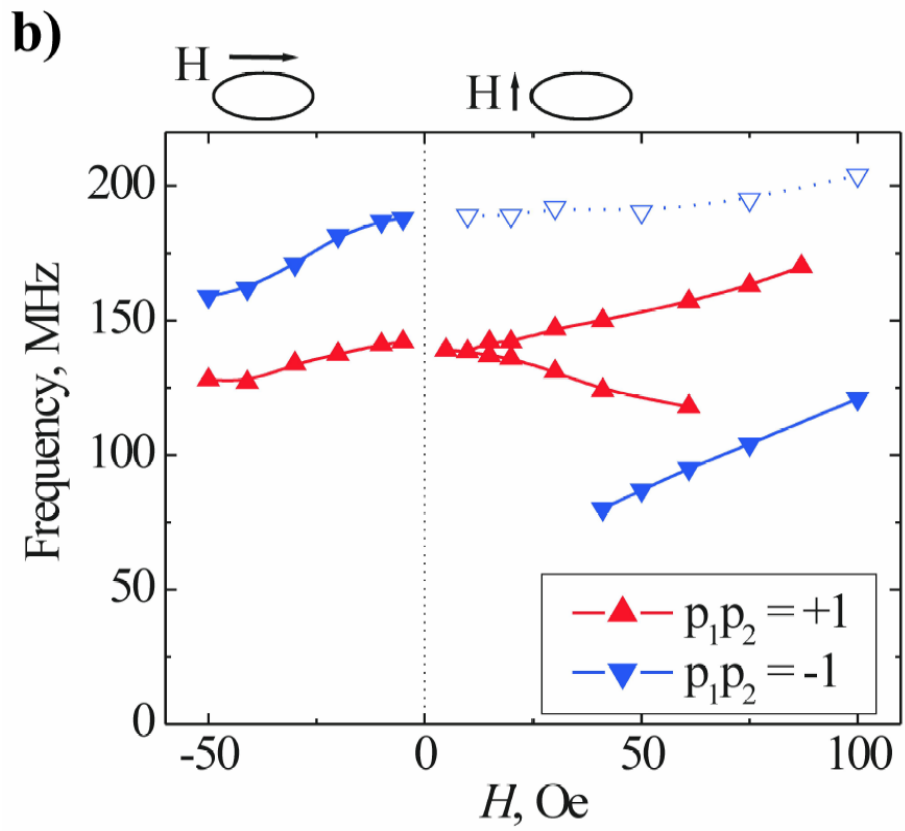

Figure 2

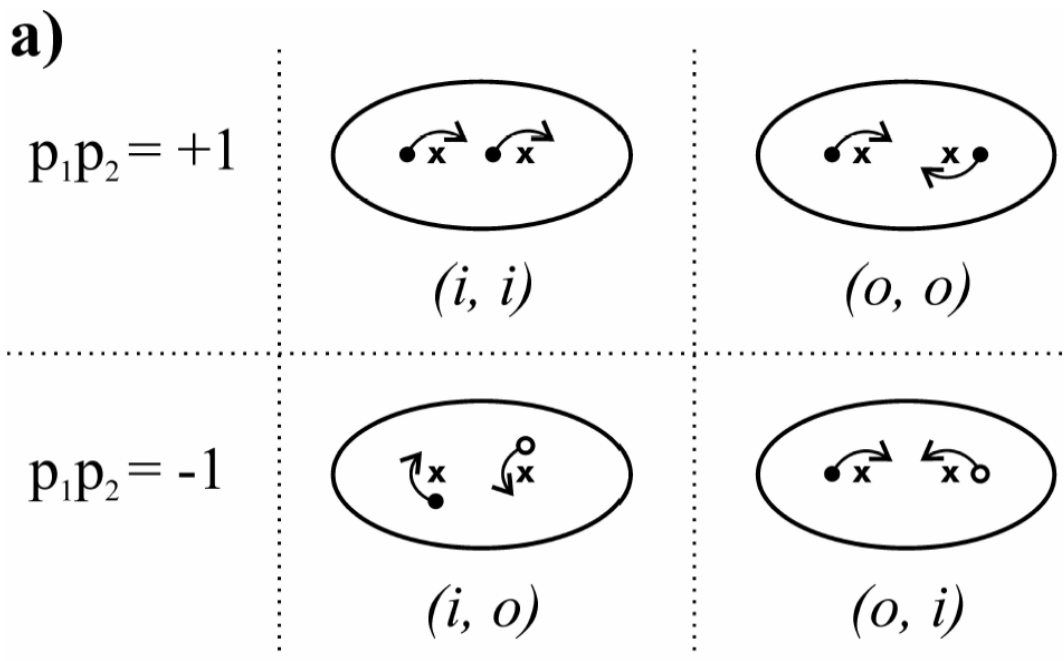

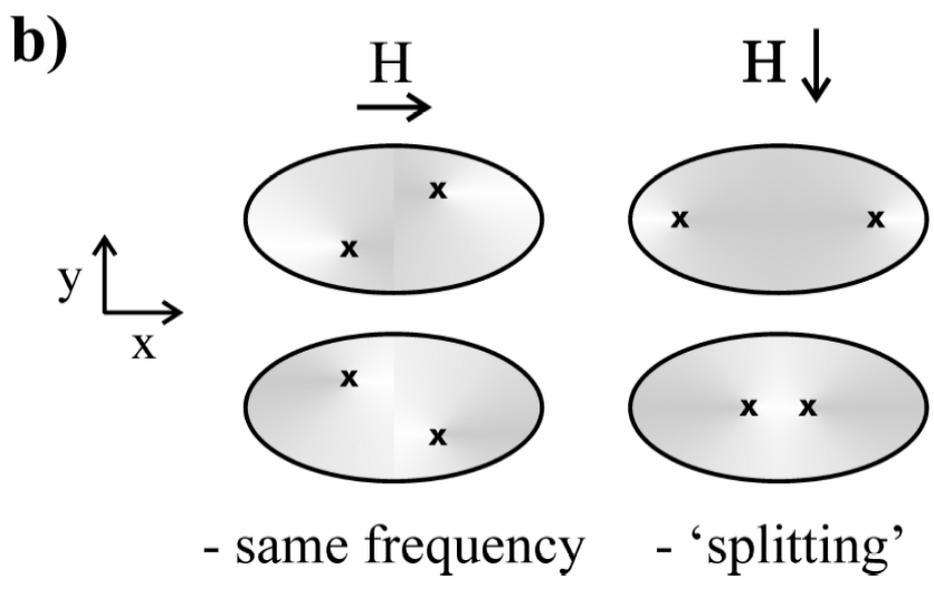

Figure 3

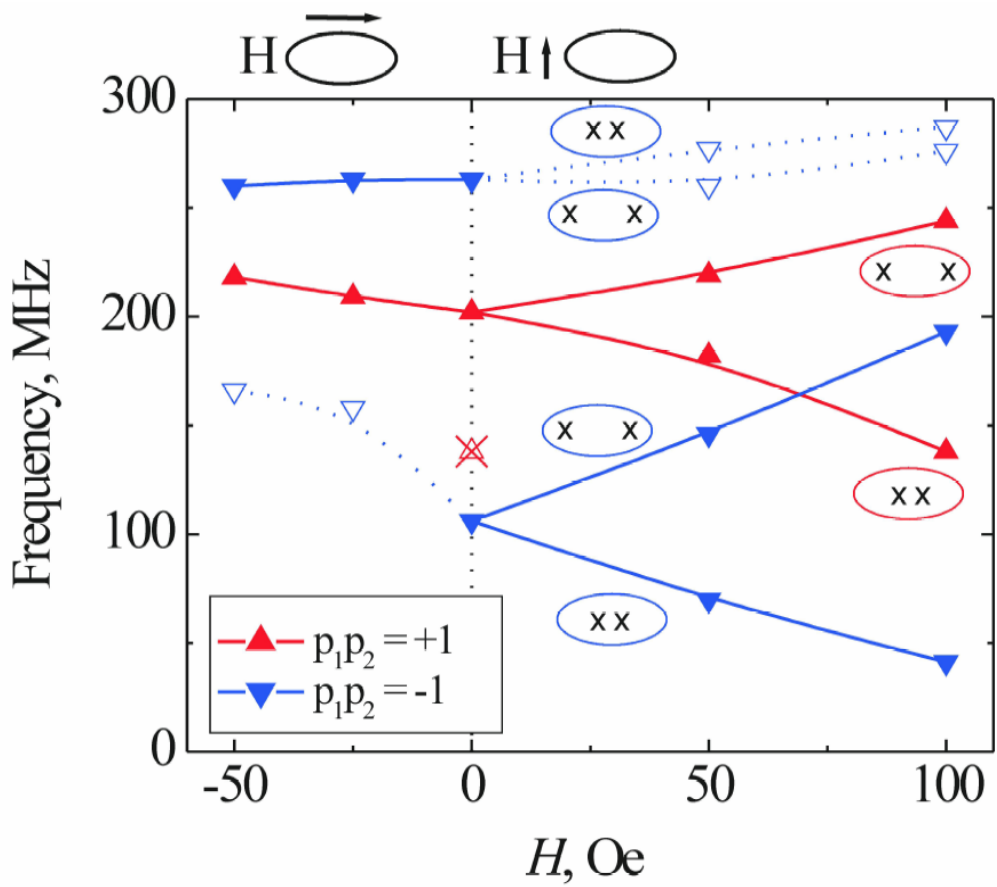

Figure 4